\documentclass[fleqn,usenatbib,usedcolumn]{mnras}
\usepackage[british]{babel}
\usepackage{newtxtext,newtxmath}
\usepackage[T1]{fontenc}  
\usepackage{graphicx}
\usepackage{float}
\usepackage{amsmath}
\usepackage{booktabs}
\usepackage{pdflscape}
\usepackage{ae,aecompl}
\usepackage{subcaption}
\usepackage{cleveref}
\usepackage{times}

\title[BHAR, SFR, and Quenching in MW and M31-Mass Progenitors]{Decoupled Black Hole Accretion and Quenching: \\ The Relationship Between BHAR, SFR, and Quenching in Milky Way and Andromeda-mass Progenitors Since $z=2.5$}

\author[M. J. Cowley et al.]{M. J. Cowley$^{1,2,3}$\thanks{E-mail: michael.cowley@students.mq.edu.au},
L. R. Spitler$^{1,2,3}$,
R. F. Quadri$^{4}$, 
A. D. Goulding$^{5}$,
C. Papovich$^{4}$,
\newauthor 
K. V. H. Tran$^{4}$,
I. Labb\'e$^{6}$,
L. Alcorn$^{4}$,
R. J. Allen$^{3,7}$,
B. Forrest$^{4}$,
K. Glazebrook$^{7}$,
\newauthor
G. G. Kacprzak$^{7}$,
G. Morrison$^{8}$,
T. Nanayakkara$^{7}$,
C. M. S. Straatman$^{9}$,
\newauthor and
A. R. Tomczak$^{10}$
\\
\\
$^{1}$Department of Physics and Astronomy, Macquarie University, NSW 2109, Australia\\
$^{2}$Research Centre for Astronomy, Astrophysics \& Astrophotonics, Macquarie University, Sydney, NSW 2109, Australia\\
$^{3}$Australian Astronomical Observatory, PO Box 915, North Ryde, NSW 1670, Australia\\
$^{4}$George P. and Cynthia W. Mitchell Institute for Fundamental Physics and Astronomy, Department of Physics and Astronomy, \\
Texas A\&M University, College Station, TX 77843, USA\\
$^{5}$Department of Astrophysical Sciences, Princeton University, Princeton, NJ\\
$^{6}$Leiden Observatory, Leiden University, PO Box 9513, 2300 RA Leiden, The Netherlands\\
$^{7}$Centre for Astrophysics and Supercomputing, Swinburne University, Hawthorn, VIC 3122, Australia\\
$^{8}$LBT Observatory, University of Arizona, 933 N. Cherry Ave. Tucson, AZ 85721, USA\\
$^{9}$Max Planck Institute for Astrophysics, Karl-Schwarzschild-Str. 1, Postfach 1317, D-85741 Garching, Germany\\
$^{10}$Department of Physics, University of California Davis, One Shields Avenue, Davis, CA 95616, USA\\
}

\date{Accepted XXX. Received YYY; in original form ZZZ}

\pubyear{2017}

\begin{document}
\label{firstpage}
\pagerange{\pageref{firstpage}--\pageref{lastpage}}
\maketitle

\begin{abstract}

We investigate the relationship between the black hole accretion rate (BHAR) and star-formation rate (SFR) for Milky Way (MW) and Andromeda (M31)-mass progenitors from $z = 0.2 - 2.5$. We source galaxies from the $K_{\mathrm{s}}$-band selected ZFOURGE survey, which includes multi-wavelenth data spanning $0.3-160 \mu$m. We use decomposition software to split the observed SEDs of our galaxies into their active galactic nuclei (AGN) and star-forming components, which allows us to estimate BHARs and SFRs from the infrared (IR). We perform tests to check the robustness of these estimates, including a comparison to BHARs and SFRs derived from X-ray stacking and far-IR analysis, respectively. We find as the progenitors evolve, their relative black hole-galaxy growth (i.e. their BHAR/SFR ratio) increases from low to high redshift. The MW-mass progenitors exhibit a log-log slope of $0.64 \pm 0.11$, while the M31-mass progenitors are $0.39 \pm 0.08$. This result contrasts with previous studies that find an almost flat slope when adopting X-ray/AGN-selected or mass-limited samples and is likely due to their use of a broad mixture of galaxies with different evolutionary histories. Our use of progenitor-matched samples highlights the potential importance of carefully selecting progenitors when searching for evolutionary relationships between BHAR/SFRs. Additionally, our finding that BHAR/SFR ratios do not track the rate at which progenitors quench casts doubts over the idea that the suppression of star-formation is predominantly driven by luminous AGN feedback (i.e. high BHARs).

\end{abstract}


\section{Introduction}

The Milky Way (MW) and Andromeda (M31) have long provided astronomers with invaluable insight in to galaxy evolution \citep{Freeman:2002kz}. With deep new surveys, it is now possible to search for their probable progenitors at high-$z$ and learn about their evolutionary history that led to their present-day properties. For example, \citet{VanDokkum:2013fg} used data from the 3D-HST \citep{Brammer:2012bu} and CANDELS \citep{Grogin:2011hx, Koekemoer:2011br} surveys to probe MW-mass progenitors out to $z = 2.5$, while \citet{Papovich:2015kn} used data from the ZFOURGE survey \citep{Straatman:2016tm} to probe MW- and M31-mass progenitors, while pushing to higher redshifts ($z \sim 3$). By investigating the evolution of various physical parameters, including rest-frame colours, morphologies, gas fractions, size, and star-formation rates, these studies point to a scenario in which the progenitors of MW- and M31-mass galaxies gradually transition from gas-rich, star-forming galaxies at high-$z$ to quenched, bulge-dominated galaxies at low-$z$. 

The growth of bulges is often associated with supermassive black holes (SMBHs) \citep[e.g.][]{Cisternas:2011ce}, where local scaling relations, manifested in the $M-\sigma$ relation \citep[e.g.][]{Marconi:2004km, McConnell:2013fq}, have driven a wealth of research in the area of SMBH-galaxy co-evolution. For example, theoretical simulations commonly invoke feedback from the active nucleus of SMBHs to regulate the star formation activity of galaxies \citep[e.g.][]{Silk:1998up, Croton:2006ew}, while several observational studies have identified correlations between AGN luminoisty and SFR \citep{Rafferty:2011bz, Chen:2013cv, Dai:2015vta, Heinis:2016tn}. However, what can be learned from these works is limited due to progenitor bias \citep[e.g.][]{VanDokkum:2001dl, Leja:2013he}. Given this, it is pertinent to ask whether the SMBH-galaxy correlations hold within the framework of evolving MW- and M31-mass progenitors, which may provide greater insight in to the processes that drive the transition of star-forming galaxies in to quiescent ones. 

In this paper, we investigate the SMBH-galaxy co-evolution of MW- and M31-mass progenitors by tracking their mean black hole accretion rates (BHARs) and mean star-formation rate (SFRs) since $z = 2.5$. We place these results in the context of galaxy quenching by comparing the evolution of the BHARs and SFRs to the quenching rate over similar timescales. This work will help provide greater insight in to the formation processes of MW- and M31-mass progenitors and to what extent the feedback from SMBH accretion plays in the quenching of galaxies over cosmic time. Throughout this paper, we use an AB magnitude system, a \citet{Chabrier:2003ki} IMF, and assume a $\Lambda$CDM cosmology with ${H_0}$ = 70 km s$^{-1}$ Mpc$^{-1}$, $\Omega_M$ = 0.3, $\Omega_{\Lambda}$ = 0.7.

\begin{figure}
\begin{center}
\includegraphics[width=\columnwidth]{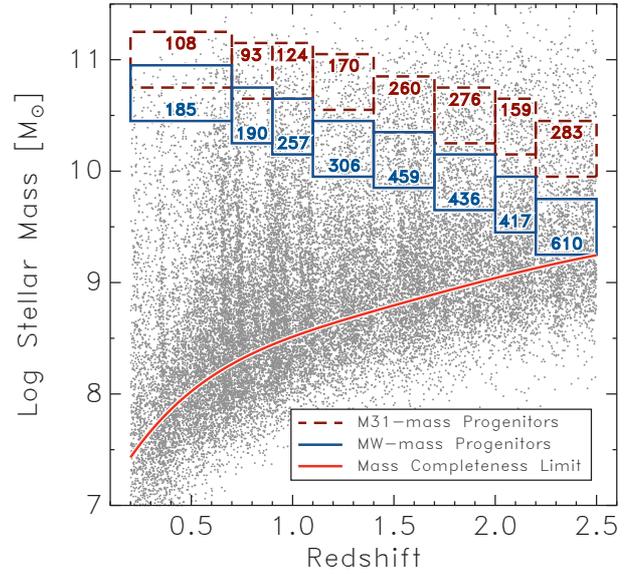}
\caption{The stellar-mass evolution of MW-mass (solid blue line boxes) and M31-mass (dashed red line boxes) galaxy progenitors, including counts for each redshift bin. The data points show the stellar masses of all sources in ZFOURGE over $z=0.2-2.5$. The red curve shows the 80 per cent stellar mass completeness limit for star forming and passive galaxies in ZFOURGE ({\color{blue}Spitler et al.} {\color{blue}2017, in prep}).}
\label{fig:fig1}
\end{center}
\end{figure}

\section{Data Sets}
\label{sec:data}

In this paper, we select galaxies from the recent public release of ZFOURGE\footnote{\url{http://zfourge.tamu.edu}} \citep{Straatman:2016tm}, with coverage in three $11' \times 11'$ pointings in the CDFS \citep{Giacconi:2002ef}, COSMOS \citep{Scoville:2007dl}, and UDS \citep{Lawrence:2007hu} fields. ZFOURGE uniquely employs deep near-IR imaging taken with six medium-band filters ($J_1, J_2, J_3$, $H_{\mathrm{s}}$, $H_{\mathrm{l}}$, $K_{\mathrm{s}}$) in FourStar \citep{Persson:2013eo} mounted on the 6.5m Magellan Baade telescope. The ultra-deep $K_{\mathrm{s}}$ detection images reach 5$\sigma$ point-source limiting depths of $\sim$26 AB mag. Near-IR imaging is supplemented with existing data from CANDELS HST \citep{Grogin:2011hx, Koekemoer:2011br, Skelton:2014do}, {\it Spitzer}/IRAC and {\it Herschel}/PACS, as well as other ground-based filters, to generate multi-wavelength catalogues spanning $0.3-160 \mu$m. Photometric redshifts were calculated in \texttt{EAZY} \citep{Brammer:2008gn} using five templates generated from the \texttt{P\'{E}GASE} library \citep{Fioc:1997up}, plus three additional dust-reddened templates \citep{Brammer:2008gn}, an passive red galaxy template \citep{Whitaker:2011cn}, and a strong emission line galaxy template \citep{Erb:2010iy}. Stellar masses were calculated by fitting the \citet{Bruzual:2003ck} models to \texttt{FAST} \citep{Kriek:2009cs}, assuming solar metallicity, a \citet{Calzetti:2000iy} dust extinction law (with $A_V$ = $0-4$), a \citet{Chabrier:2003ki} initial mass function (IMF) and exponentially declining star-formation histories of the form SFR($t$) $\propto {e}^{{-t}/{\tau}}$.

To estimate the average BHAR of the MW- and M31-mass progenitors, we measure AGN luminosities using a combination of IR and X-ray observations. We make use of overlapping {\it Spitzer}/MIPS and {\it Herschel}/PACS far-infrared (FIR) imaging, which is sourced from the GOODS {\it Spitzer} Legacy program (PI: M. Dickinson) and GOODS-H \citep{Elbaz:2011ix} for the ZFOURGE-CDFS field, S-COSMOS Spitzer Legacy program (PI: D. Sanders) and CANDELS-H ({\color{blue}Inami et al.} {\color{blue}2017, in prep}) for the ZFOURGE-COSMOS field, and SpUDS {\it Spitzer} Legacy program (PI: J. Dunlop) and CANDELS-H for the ZFOURGE-UDS field. For X-ray observations, we make use of the deepest {\it Chandra} imaging available, which is sourced from the {\it Chandra} Deep Field-South Survey: 7 Ms Source catalogs \citep{Luo:2017be} for the ZFOURGE-CDFS field, the {\it Chandra} COSMOS Survey I. Overview and Point Source catalogue \citep{Elvis:2009io} for the ZFOURGE-COSMOS field, and the X-UDS Chandra Legacy Survey (PI: Hasinger) for the ZFOURGE-UDS field.

\begin{figure*}
\begin{center}
\includegraphics[width=\textwidth]{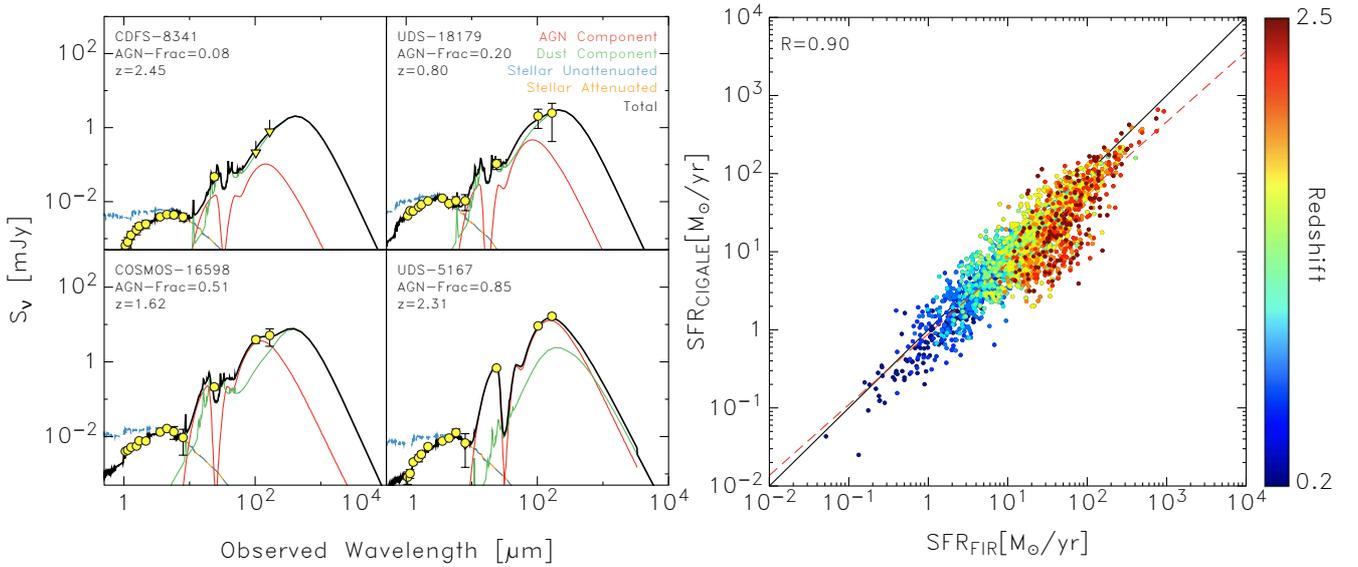}
\caption{{\it Left panels:} SED decomposition on a selection of ZFOURGE sources using \texttt{CIGALE} \citep{Burgarella:2005bd, Noll:2009ed}. Yellow circles are the observed points, yellow arrows are the upper limits, and the black lines are the best-fit total models. We also show the AGN component in solid red lines, stellar-heated dust component in solid green lines, the unattenuated stellar emission in dashed blue lines, and the attenuated stellar emission in solid orange lines. The corresponding redshift and fraction of AGN emission to the $L_{\mathrm{IR}}$ for each source are also provided. {\it Right panels:} a comparison of \texttt{CIGALE}-derived SFRs to those from a FIR-derived conversion of the bolometric 8-1000$\mu$m IR luminosity calculated from a luminosity-independent conversion \citep{Wuyts:2008hi, Wuyts:2011da} using PACS 160$\mu$m fluxes. The solid black line is the 1:1 relation, while the dashed red line is the best fit. We also show the linear Pearson correlation coefficient, R.}
\label{fig:fig2}
\end{center}
\end{figure*}

\section{Data Analysis}
\label{sec:analysis}
\subsection{Progenitor Selection}
\label{sec:selectprog}

To investigate the evolution of MW- and M31-mass galaxies, we select progenitors with present-day stellar masses near those of the MW \citep[$M_* = 5 \times 10^{10} M_{\odot}$ at $z = 0$; ][]{McMillan:2011hf, VanDokkum:2013fg, Licquia:2015eb} and M31 \citep[$M_* = 10^{11} M_{\odot}$ at $z = 0$; ][]{Mutch:2011dr}. Progenitor galaxies were selected using the approach in \citet{Papovich:2015kn}, who traced the stellar-mass evolution of present-day MW- and M31-mass galaxies using the multi-epoch abundance matching (MEAM) method of \citet{Moster:2013dl}. From this work, \citet{Moster:2013dl} derived the fitting functions for the star formation history and mass accretion history for galaxies of arbitrary present-day stellar mass. \citet{Papovich:2015kn} then integrated the fitting functions with respect to time, accounting for mass losses from stellar evolution, to derive the stellar mass evolution of the present day MW-mass and M31-mass galaxies. As shown in Figure~\ref{fig:fig1}, estimates for the 80 per cent mass completeness limits mean the data from ZFOURGE is unlikely to introduce selection biases in our attempt to track the stellar mass evolution of progenitors to $z=2.5$. We identify 2,860 MW-mass galaxy progenitors and 1,473 M31-mass galaxy progenitors, spanning $z=0.2-2.5$.

\subsection{Black Hole Accretion Rates}

The luminosity emitted by an AGN is a result of a mass-accretion event \citep[e.g.][]{Alexander:2012er} which can be described by $L_{\mathrm{AGN}} = \epsilon c^2 dM/dt$, where $\epsilon$ is the accretion efficiency \citep[often estimated to be $\epsilon = 0.1$; e.g.][]{Marconi:2004km}. In units of $M_{\odot}\mathrm{yr}^{-1}$, the black hole accretion rate (BHAR) can be expressed as:

\begin{equation}
\label{eq:1}
\mathrm{BHAR} = 0.15\frac { \epsilon  }{ 0.1 } \frac { { L }_{ { \mathrm{AGN} } } }{ { 10 }^{ 45 }\mathrm{erg\:s}^{ -1 } }
\end{equation}

\noindent where $L_{\mathrm{AGN}}$ is the AGN bolometric luminosity. In the following section, we describe the methods used to estimate $L_{\mathrm{AGN}}$ for all AGN in our sample.

\subsection{SED Decomposition of $L_{\mathrm{IR}}$}
\label{sec:decomp}

We use the multi-component SED fitting code, \texttt{CIGALE}\footnote{\url{http://cigale.lam.fr}} \citep[Code Investigating GALaxy Emission;][]{Burgarella:2005bd, Noll:2009ed} to decompose the rest-frame IR luminosity ($L_{\mathrm{IR}}$) of MW- and M31-mass galaxy progenitors in to their AGN and star-forming components. By binning these components, we respectively estimate the mean BHARs and SFRs of the progenitors in bins of stellar mass and redshift. In Table~\ref{table:params}, we list the parameters we use to complete SED fitting and decomposition. \texttt{CIGALE} completes decomposition using a two step process. First, it creates a library of SED models using the chosen parameters, before identifying the best-fit model to the observed photometry through $\chi^2$ minimisation. Galaxy parameters and their associated uncertainties are estimated using a Bayesian approach, which derives the probability that each parameter value is representative of a given galaxy \citep[see][]{Burgarella:2005bd}. From these various parameters, we focus on the recovered $L_{\mathrm{IR}}$, which consists of contributions from stellar-heated dust (dominated by young stars) and AGN-heated dust ($L_{\mathrm{AGN}}$). 

For the stellar-heated dust, we adopt the semi-empirical templates of \citet{Dale:2014fb}, which include modified templates from the \citet{Dale:2002bo} library. For the AGN-heated dust, we adopt the templates of \citet{Fritz:2006eo}, which consider the emission of the central source as well as the radiation from the dusty torus. The \citet{Fritz:2006eo} templates introduce six additional parameters (see Table~\ref{table:params}), which describe the geometrical configuration of the torus and the properties of the dust emission. We fix these parameters to mean values based on studies that extensively test AGN fitting with \texttt{CIGALE} \citep{Ciesla:2015vi, Heinis:2016tn, Bernhard:2016tl, Wylezalek:2016ik}. By fixing these parameters, we reduce the parameter space and hence the overall degeneracy of the models, without compromising the recovery of the components. Further details are available in \citet{Ciesla:2015vi}, while Table~\ref{table:params} lists the parameters we use to complete SED fitting and decomposition.

A caveat to this approach is \texttt{CIGALE}'s reliability at low AGN luminosities. \citet{Ciesla:2015vi} found for such sources that the software would tend to overestimate the AGN contribution up to $\sim 120\%$. The authors attribute this overestimation to bias from the PDF analysis, where the PDF is truncated and returns an elevated value. To address this, we select all sources with error\_$L_{\mathrm{AGN}} > L_{\mathrm{AGN}}$ and scatter them down by randomly drawing a new $L_{\mathrm{AGN}}$ value from a gaussian centred on zero with a standard deviation of error\_$L_{\mathrm{AGN}}$. We compute the averages reported below using these new $L_{\mathrm{AGN}}$ values whenever \texttt{CIGALE} returns a non-detection of an IR AGN component. For a secure detection, we adopt the output directly from \texttt{CIGALE}

\begin{figure}
\begin{center}
\includegraphics[width=\columnwidth]{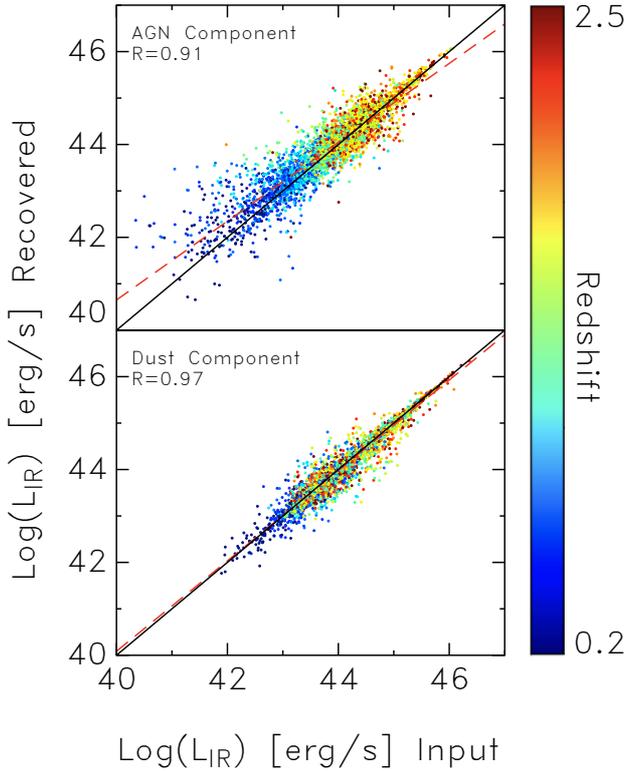}
\caption{Comparison of the AGN-heated (top panel) and stellar-heated dust components (bottom panel) recovered by \texttt{CIGALE} for our mock galaxy SEDs. Points are coloured by redshift. The solid black line is the 1:1 relation, while the dashed red line is the best fit. We also show the linear Pearson correlation coefficient, R.}
\label{fig:fig3}
\end{center}
\end{figure}

Previous studies, which have used \texttt{CIGALE} to decompose the $L_{\mathrm{IR}}$ have shown robust luminosity estimations are heavily reliant on rest-frame IR data \citep{Buat:2013bl, Ciesla:2015vi}. Therefore, we use FourStar (1.1, 1.2, 1.3, 1.6, 1.7, and 2.2$\mu$m), IRAC (3.6, 4.5, 5.8, and 8$\mu$m), MIPS (24$\mu$m), and PACS (100 and 160$\mu$m) broadband data in our SED fitting. While all sources are detected in multiple bands, only $\sim$65 per cent are detected in at least one near-IR, one mid-IR, and one far-IR band. For non-detections (flux < 0), we replace flux values with their corresponding uncertainties and treat them as upper limits. Examples of best-fit models and decomposition are shown in the left panel of Figure~\ref{fig:fig2}.

We apply two tests in order to assess CIGALE's ability to robustly estimate parameters. The first is a check of the \texttt{CIGALE}-derived SFRs, which we achieve by comparing the results to those from a FIR-derived conversion of the bolometric 8-1000$\mu$m IR luminosity calculated from a luminosity-independent conversion \citep{Wuyts:2008hi, Wuyts:2011da} using PACS 160$\mu$m fluxes. The results, which are presented in the right panel of Figure~\ref{fig:fig2}, show a strong correlation between the two methods of derivation. A noticeable exception is for a selection of high redshift sources, which stray from the 1:1 line. When we investigate these sources, we find they are dominated by AGN (i.e. greater than 50\% AGN-heated component) to the $L_{\mathrm{IR}}$. While the FIR regime is believed to be largely uncontaminated by AGN \citep{Netzer:2007bs, Mullaney:2012ce}, it is not completely immune from AGN-dominated sources towards higher redshifts \citep[see][]{Cowley:2016dv}. This is likely why some of our AGN-dominated sources at high redshift exhibit FIR-derived SFRs that are elevated over their \texttt{CIGALE}-derived counterparts.

The second test we perform is by way of \texttt{CIGALE}'s mock utility, which generates a mock catalogue of artificial SEDs using the best-fit templates to the observed SEDs. The mock catalogue is built by integrating the best-fit SED of each source in the observed bands, before random noise, distributed assuming Gaussian errors with the observed error as the standard deviation, is added to the fluxes. We then run \texttt{CIGALE} on the mock galaxy SEDs and compare the input parameters to the recovered parameters. The results are shown in Figure~\ref{fig:fig3}. For both the AGN-heated and stellar-heated dust components, we find very good correlation with R$>0.90$, suggesting \texttt{CIGALE}'s ability to recover parameters is robust, despite the limited filter set and typical flux errors of our observational data. We also point the reader to \citet{Ciesla:2015vi}, for a detailed study of broadband SED fitting methods and the reliability of \texttt{CIGALE} to recover parameters via decomposition.

\begin{table}
\setlength{\tabcolsep}{0.45em}
\centering
\caption{Modules and Parameters used in \texttt{CIGALE}}
\begin{tabular}{l c}
\hline
\noalign{\vskip 1mm} 
{\bf Module} & {\bf Model} \\ [0.5ex]
\hline  
\noalign{\vskip 1mm} 
star-formation history & Delayed ${\tau}$ \\[1ex]
Single Stellar Population models & \citet{Bruzual:2003ck} \\[1ex]
Initial Mass Function & \citet{Chabrier:2003ki} \\[1ex]
Attenuation law & \citet{Calzetti:2000iy} \\[1ex]
Dust emission models & \citet{Dale:2014fb} \\[1ex]
AGN emission models & \citet{Fritz:2006eo} \\[1ex]
\hline
\noalign{\vskip 1mm} 
{\bf Parameter} & {\bf Value} \\ [0.5ex]
\hline  
\noalign{\vskip 1mm}  
E-folding timescale$^1$, $\tau$ (Gyr) & 1, 3, 5, 7, 9, 11 \\[1ex]
Age of oldest stars$^1$, $t$ (Gyr) & 1, 3, 5, 7, 9, 11 \\[1ex]
E(B-V)$_*$ for young population & 0.01, 0.05, 0.1, 0.5, 1.0, 1.5 \\[1ex]
Ratio of torus radii$^2$ & 60 \\[1ex]
Optical depth at 9.7$\mu$m of torus$^2$ & 0.3, 3.0, 6.0, 10.0 \\[1ex]
Parameter for torus density$^{2,3}$, $\beta$ & -0.5 \\[1ex]
Parameter for torus density$^{2,3}$, $\gamma$ & 0 \\[1ex]
Opening angle of torus$^2$ & 100 \\[1ex]
Angle of AGN axis to line of sight$^2$ & 0.001, 50.100, 89.990 \\[1ex]
AGN fraction of $L_{\mathrm{IR}}$$^2$ & 0.00 - 0.95 (steps of 0.05) \\[1ex]
\hline
\end{tabular}
  \\[0.2cm] 
  \raggedright \footnotesize $^1$ SFR($t$) $\propto {e}^{{-t}/{\tau}}$\\
  						     $^2$ AGN parameters from \citet{Fritz:2006eo}\\
  							 $^3$ $\rho$($r, \theta$) = $\alpha r^\beta $e$^{-\gamma\left| \mathrm{cos}(\theta) \right|}$
\label{table:params}
\end{table}

\subsection{$L_{\mathrm{AGN}}$ from X-ray Stacking}

We use the X-ray stacking code, \texttt{STACKFAST}\footnote{\url{http://www.dartmouth.edu/~stackfast/}} \citep{Hickox:2007fb} to estimate the average X-ray luminosity for MW- and M31-mass galaxy progenitors in bins of redshift. X-ray stacking allows us to account for those sources that are not individually detected in X-rays. We determine the stacked source count rate (in counts s$^{-1}$) for the M31- and MW-mass progenitors in bins of redshift and convert to flux (in ergs cm$^{-2}$ s$^{-1}$) in the 0.5-7 keV band assuming an intrinsic X-ray spectrum with $\Gamma = 1.8$. 
We derive the average X-ray luminosity using:

\begin{equation}
\quad { L }_{ \mathrm{X} }[\mathrm{erg \ s}^{-1}\mathrm{]} =4\pi { d }_{ l }^{ 2 }{ \left( 1+z \right)  }^{ \Gamma -2 }{ f }_{ x }
\end{equation}

\noindent where ${ d }_{ l }$ is the average luminosity distance determined for each redshift bin. Finally, to convert $L_{\mathrm{x}}$ to $L_{\mathrm{AGN}}$, we apply a constant bolometric correction factor of 22.4 (based on a sample of local, $L_{\mathrm{x}}$ = 10$^{41-46}$ erg s$^{-1}$, AGNs from \citealt{Vasudevan:2007ew}). More details of \texttt{STACKFAST} are described in Section 5.1 of \citet{Hickox:2007fb}, while the basics of our X-ray data processing, reduction and image analysis can be found in \citet{Goulding:2012cn}. 

\section{Results}

\subsection{Evolution of Star-Formation and Black Hole Accretion}

\subsubsection{Evolution of the Black Hole Accretion Rates}

Using Equation (\ref{eq:1}), we estimate the mean BHAR of all galaxies in bins of stellar mass and redshift and plot the BHAR history of the MW- and M31-mass progenitors in Figure~\ref{fig:fig4} (top panel). Both the X-ray and IR-derived BHARs start relatively high in the highest redshift bins and track a similar path as they reduce in rate towards the present day. The notable difference between the the two samples is a significant offset. Specifically, we find the IR-derived BHARs to be $\sim$4 times higher than the X-ray BHARs. One likely cause of this discrepancy is absorption effects. Currently, we assume no absorption during the X-ray analysis, but if we let the average intrinsic neutral hydrogen column density for the X-ray sample be $N_{\mathrm{H}} \sim 3 \times 10^{23}$ cm$^{-2}$ (i.e. heavily obscured), this would fully account for the offset. Such levels of obscuration are supported by the flux hardness ratios (HR) of our X-rays\footnote{We calculate flux hardness ratio between the 0.5-2 and 2-7 keV bands. We find HR values consistent with moderate to heavy obscuration, with the M31-mass progenitors systematically higher (HR $\sim$ 4) than the MW-mass progenitors (HR $\sim$ 3.8)} A similar elevation for IR-derived BHARs was found in \cite{Gruppioni:2011ei}.

\subsubsection{Evolution of the Star-Formation Rates}

Figure~\ref{fig:fig4} (middle panel) shows the evolution of the mean SFRs, in bins of stellar mass and redshift, for the progenitors. For the M31-mass progenitors, SFRs start high ($> 30 \: M_{\odot}$ yr$^{-1}$) at the highest redshifts observed, before peaking at $\sim 40 \: M_{\odot}$ yr$^{-1}$ around $z \sim 1.75$. Following this peak, the SFRs for the M31-mass progenitors decline monotonically to values of a few solar masses per year at z = 0.2. The MW-mass progenitors follow a similar trend, but are lower at $z > 1$. The SFRs for MW-mass progenitors start at $\sim 5 \: M_{\odot}$ yr$^{-1}$ in the highest redshift bins, peak at $\sim 15 \: M_{\odot}$ yr$^{-1}$ around $z \sim 1.5$, and then decline at similar values to the M31-mass progenitors at $z<1$. The evolution of the mean SFRs in Figure~\ref{fig:fig4} are found to qualitatively match those in \citet{VanDokkum:2013fg} and \citet{Papovich:2015kn}, albeit slightly lower in value. As mentioned in Section~\ref{sec:decomp}, this offset is likely attributed to the different approach to deriving SFRs and the removal of AGN emission performed in this study.

\subsubsection{Evolution of the Relative Black Hole-Galaxy Growth}

Figure~\ref{fig:fig4} (bottom-panel) shows the evolution of the ratio between the BHAR and SFR for the progenitors. In all cases, we find the BHAR/SFR ratios increase with redshift for the MW- and M31-mass progenitors. Upon applying a least-squares fit, we find the slopes of the MW-mass progenitors to be $0.64(\pm 0.11)$ (i.e. log[BHAR/SFR] $=0.64 (\pm 0.11) \times z -3.52$) and $0.55(\pm 0.10)$ for the IR-derived and X-ray-derived BHAR/SFR ratios, respectively. This is marginally stronger than the M31-mass progenitors, which exhibit slopes of $0.39 (\pm 0.08)$ and $0.08(\pm 0.08)$. The flatter slope of the massive M31-mass progenitors is more consistent with studies that adopt different sample selection, such as \citet{Calhau:2017if} who find an almost flat relationship of $\sim 10^{-3.2}$ in BHAR/SFR over $z=0-2.23$ for H$\alpha$-selected star-forming galaxies.

\subsubsection{Evolution of the Quiescent Fraction and Quenching Rate} 

\begin{figure}
\begin{center}
\includegraphics[width=\columnwidth]{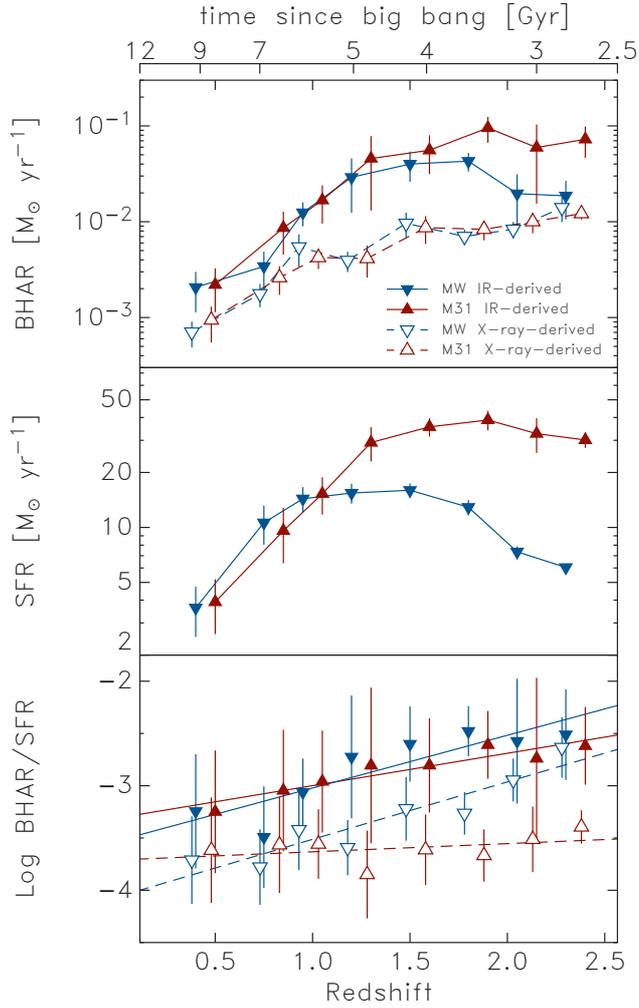}
\caption{{\it Top panel:} The mean BHAR as a function of redshift for our MW- (blue; down triangles) and M31-mass (red; up triangles) progenitors. Vertical errors represent errors on the mean. We apply a slight offset in redshift for clarity. {\it Middle panel:} the mean SFR as a function of redshift for our progenitors (same symbols as top panel). {\it Bottom panel:} the mean BHAR to SFR ratio as a function of redshift for our progenitors (same symbols as top panel). The solid lines indicate a least-squares linear fit to these data.}
\label{fig:fig4}
\end{center}
\end{figure}

\begin{figure}
\begin{center}
\includegraphics[width=\columnwidth]{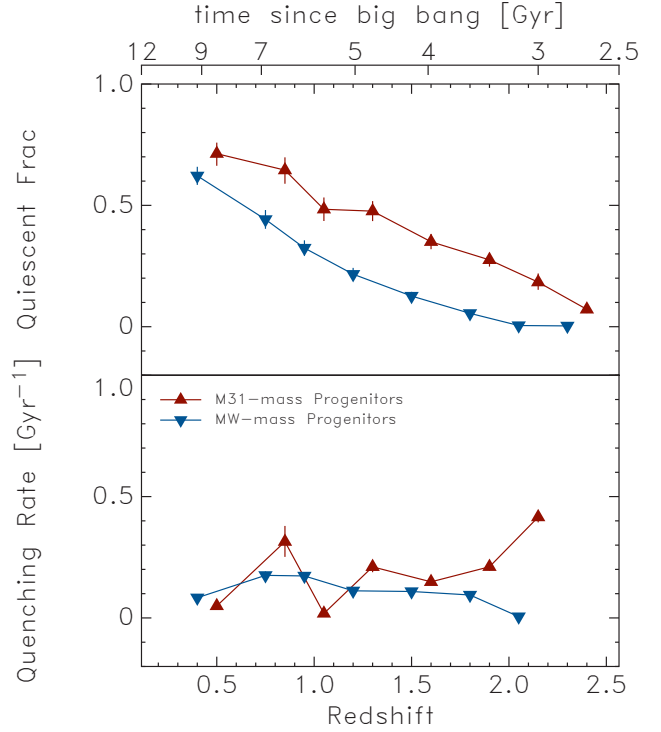}
\caption{{\it Top panel:} The evolution of the quiescent fraction ($f_{\mathrm{quies}}=N_{\mathrm{quies}}/(N_{\mathrm{quies}}+N_{\mathrm{sf}})$) for our MW- (blue; down triangles) and M31-mass (red; up triangles) progenitors as a function of redshift. We apply a slight offset in redshift for clarity. Vertical errors represent the binomial confidence interval. {\it Bottom panel:} the evolution of the quenching rate ($( { f }^{ \mathrm{sf} }_{ \mathrm{zbin(n)} }-{ f }^{ \mathrm{sf} }_{ \mathrm{zbin(n-1)} } ) /{ f }^{ \mathrm{sf} }_{ \mathrm{zbin(n)} }/{ \mathrm{Gyr} }_{ \mathrm{zbin(n-1)-zbin(n)} }$) for our progenitors (same symbols as top panel).}
\label{fig:fig5}
\end{center}
\end{figure}

Figure~\ref{fig:fig5} (top-panel) shows the evolution of the quiescent fraction of the progenitors, where the quiescent fraction is defined as the ratio of the total number of quiescent galaxies to the total number of galaxies ($f_{\mathrm{quies}}=N_{\mathrm{quies}}/(N_{\mathrm{quies}}+N_{\mathrm{sf}})$). We seperate quiescent galaxies from star-forming galaxies using $UVJ$-colour analysis \citep[see][]{Cowley:2016dv}. Errors are calculated using the Clopper-Pearson approximation of the binomial confidence interval. For both samples, the quiescent fraction increases with decreasing redshift. We also show the quenching rate (bottom-panel), which is the rate at which the progenitors quench (i.e. move from star-forming to quiescent in $UVJ$-colour space) per gigayear. We quantify the quenching rate as the probability that a star-forming progenitor will become quenched per unit time, i.e $( { f }^{ \mathrm{sf} }_{ \mathrm{zbin(n)} }-{ f }^{ \mathrm{sf} }_{ \mathrm{zbin(n-1)} } ) /{ f }^{ \mathrm{sf} }_{ \mathrm{zbin(n)} }/{ \mathrm{Gyr} }_{ \mathrm{zbin(n-1)-zbin(n)} }$. Evidence for AGN quenching would likely return high quenching rates during periods of high BHAR. Instead, we find the evolution of the BHAR for the the MW- and M31-mass progenitors (see Figure~\ref{fig:fig4}) to be decoupled from the quenching rate over similar timescales.

\section{Discussion}

Tight correlations between BHAR and SFR are well documented \citep[e.g.][]{Merloni:2008kc, Aird:2010ko, Gruppioni:2011ei, Delvecchio:2014gp} with both models \citep{Silk:2013eo} and observations \citep{Calhau:2017if, Mullaney:2012df, Dai:2015vta} producing a nearly flat BHAR/SFR ratio across cosmic time. This flat correlation is often explained by a simple scenario where a joint fuelling process regulates both SMBH growth and star-formation \citep[see][]{Mullaney:2012df}. 

In contrast to past work, our results show that the BHAR/SFR ratios of our progenitor samples tend to decrease towards the present day. As this appears to hold whether we use X-ray or IR-derived BHARs (see the caveat for M31 below), we hypothesise this difference is driven by sample differences as previous efforts have used various sample galaxy selections (e.g. X-ray or mass-limited samples). To test this hypothesis, we limited our sample to X-ray selected AGN in ZFOURGE \citep{Cowley:2016dv} and examined the evolution of the BHAR/SFR ratios for the MW- and M31-mass progenitors using both X-ray and IR-derived BHARs. We find an almost completely flat relationship in BHAR/SFR ratios (slopes of $-0.04$ and $-0.11$, respectively) across all redshifts, which is consistent with the literature \citep[e.g.][]{Stanley:2015eq}.

While we find an evolving BHAR/SFR ratio for the bulk of our sample, a possible exception is the M31-mass progenitors, which exhibit an almost flat ratio when using the X-ray-derived BHARs. While this differs from the IR-derived BHAR/SFR ratios of the M31-mass progenitors, we postulate this may be driven by obscuration effects, where the X-rays of the more massive, high-redshift M31-mass progenitors are highly-obscured \citep[e.g.][]{Polletta:2008cr, Treister:2008jp}. Indeed, when we investigate the X-ray hardness ratios for the M31-mass progenitors, we find results consistent with heavy obscuration (HR $\sim$ 4) at this redshift range. Therefore, such results argue for the inclusion of IR-based AGN whenever possible to fully assess the impact of dust obscuration changes.
 
The apparent differences we find for the evolution of BHAR/SFR ratios compared to past work illustrates the importance of selecting progenitors samples when looking for evolutionary changes in AGN. Indeed, mass-limited, star-forming or X-ray selected may not capture underlying evolutionary trends due to the fact that these samples contain galaxies with very different evolutionary paths \citep{Leja:2013he}. The present work directly addresses this by adopting a selection that attempts to account for the mass growth of galaxies over the redshift range considered here. 

Finally, the decline of the mean BHARs and SFRs with decreasing redshift casts doubts over the suppression of star-formation being predominantly driven by luminous AGN feedback (i.e. high BHARs) in MW- and M31-mass progenitors. While one may expect to see an increase in BHARs during a period of quenching, we instead find that the rate at which the progenitors quench (see Figure~\ref{fig:fig5}) is decoupled from the BHARs, which decline over similar timescales. An alternative scenario to explain this is one of morphological quenching \citep{Martig:2009kz}, where the formation of a bulge stabilises gas in the galactic disk and suppresses the efficiency of star-formation \citep[e.g.][]{Martig:2010ju, Ceverino:2010eh, Genel:2012bj, Sales:2012eb, Genzel:2014js}. 

This scenario is supported by the work of \citet{Papovich:2015kn}, who find the S\'{e}rsic index of the same progenitors to increase with decreasing redshift, suggesting a growth in spheroid size towards the present. This view is consistent with recent findings that AGN feedback may only play a dominant roll in star-formation quenching at lower-$z$, during periods of low-level (i.e. radio-mode) activity in bulge-dominated hosts \citep[e.g.][]{Gurkan:2015wm, Cowley:2016dv}. With this said, if bulge growth remains closely tied to SMBH growth throughout cosmic time, then our BHARs suggest that bulge growth is actually decoupled from quenching.

\section*{ACKNOWLEDGMENTS}

Australian access to the Magellan Telescopes was supported through the National Collaborative Research Infrastructure Strategy of the Australian Federal Government. Additional scientific results are based in part on observations taken by the CANDELS Multi-Cycle Treasury Program with the NASA/ESA HST, operated by the Association of Universities for Research in Astronomy, Inc., under NASA contract NAS5-26555; XMM-Newton, an ESA science mission with instruments and contributions directly funded by ESA Member States and NASA;  and the {\it Chandra} X-ray Observatory. Research support to MJC and RJA is provided by the Australian Astronomical Observatory. KG acknowledges the support of the Australian Research Council through Discovery Proposal awards DP1094370, DP130101460, and DP130101667. GGK acknowledges from the Australian Research Council through the award of a Future Fellowship (FT140100933). We acknowledge support from Texas A\&M University and the George P. and Cynthia Woods Mitchell Institute for Fundamental Physics and Astronomy.

\footnotesize{

\bibliographystyle{mnras}
\bibliography{references} 
}
\label{lastpage}
\end{document}